# Serial Correlations in Single-Subject fMRI with Sub-Second TR

Saskia Bollmann[1], Alexander M. Puckett[2], Ross Cunnington[2,3], Markus Barth[1]

[1]Centre for Advanced Imaging, The University of Queensland, Brisbane QLD 4072, Australia

[2]Queensland Brain Institute, The University of Queensland, Brisbane QLD 4072, Australia

[3]School of Psychology, The University of Queensland; Brisbane QLD 4072; Australia

Corresponding author:

Saskia Bollmann

Centre for Advanced Imaging

Building 57, Research Road

The University of Queensland

St Lucia QLD 4072

Australia

saskia.bollmann@cai.uq.edu.au

## Abstract

When performing statistical analysis of single-subject fMRI data, serial correlations need to be taken into account to allow for valid inference. Otherwise, the variability in the parameter estimates might be under-estimated resulting in increased false-positive rates. Serial correlations in fMRI data are commonly characterized in terms of a first-order autoregressive (AR) process and then removed via pre-whitening. The required noise model for the pre-whitening depends on a number of parameters, particularly the repetition time (TR). Here we investigate how the sub-second temporal resolution provided by simultaneous multislice (SMS) imaging changes the noise structure in fMRI time series. We fit a higher-order AR model and then estimate the optimal AR model order for a sequence with a TR of less than 600 ms providing whole brain coverage. We show that physiological noise modelling successfully reduces the required AR model order, but remaining serial correlations necessitate an advanced noise model. We conclude that commonly used noise models, such as the AR(1) model, are inadequate for modelling serial correlations in fMRI using sub-second TRs. Rather, physiological noise modelling in combination with advanced pre-whitening schemes enable valid inference in single-subject analysis using fast fMRI sequences.

## Introduction

In functional MRI (fMRI), blood-oxygen-level-dependent (BOLD)-weighted images are acquired in a rapid, successive fashion depicting changes in deoxyhemoglobin content in the brain over time (Bandettini et al., 1992; Kwong et al., 1992; Ogawa et al., 1992). The signal of interest is the time series in each voxel comprising the local hemodynamic response (Boynton et al., 1996; Buxton et al., 1998; Fox et al., 1988). A statistical model is then applied to draw an inference about the effect of an experimental manipulation and its significance. In this process, serial correlations, i.e. correlations between the errors of successive samples, need to be taken into account for valid inference.

Statistical analysis of fMRI time series is often performed using a mass-univariate general linear model to assess the effect of a task in each voxel (Friston et al., 1994). The basic model at each voxel reads $Y = Xw + e$; where $Y$ are the observations (data), $X$ is the design matrix containing the explanatory variables, $w$ are the (unknown) parameters on which the inference is based and $e$ is the error following a normal distribution with $e \sim N(0, \sigma^2 V)$ (Kiebel and Holmes, 2007). Here, $V$ describes the serial correlations over time. When performing a statistical test, this correlation structure needs to be taken into account to obtain the desired false-positive rate, such that the estimated variance of $\hat{w}$ is not biased, and too liberal thresholds are prevented (Purdon and Weisskoff, 1998; Woolrich et al., 2001; Worsley et al., 2002). Importantly, the estimate of $w$ itself remains unbiased but becomes more variable when disregarding serial correlations in the data (Wooldridge, 2013; Worsley et al., 2002). Serial correlations initially reduce the effective degrees of freedom (Worsley and Friston, 1995), and pre-whitening[1] is traditionally performed to remove the serial correlations from the data and the model (Woolrich et al., 2001).

In combination with physiological noise modelling, pre-whitening using an autoregressive (AR) model of order 1 has been a successful strategy to improve the validity of drawn inferences for fMRI time series with a repetition time (TR) above 2 seconds (Lund et al., 2006). Similarly, Worsley et al. (2002) found an AR(1) to be adequate for most voxels. However, it has been hypothesized that a higher AR model order might be required to reliably capture serial correlations for data sets with considerably shorter TRs (Lund et al., 2006). Given the recent acceleration of echo-planar imaging (EPI) employing simultaneous multislice (SMS) techniques (Breuer et al., 2005; Larkman et al., 2001; Setsompop et al., 2012) - for a recent review see Barth et al. (2016) - one can now easily achieve sub-second temporal resolution for whole-brain acquisitions (Feinberg et al., 2010; Moeller et al., 2010). Hence, it is becoming increasingly important to address the effect of short TRs in fMRI analyses. Increased false-positive rates were observed empirically in resting-state data with a TR of 1 second but without physiological noise modelling (Eklund et al., 2012). Similarly, when using higher AR-model orders for pre-whitening in an event-related paradigm, a reduction in t-values was found for sequences with TRs below 2 seconds (Sahib et al., 2016). However, the underlying mechanisms and sources of serial correlations remain unknown. Therefore, we investigate serial correlations in fMRI time series with sub-second TR achieved using SMS EPI to assess changes in the noise correlation structure that need to be taken into account for valid inference.

We characterize serial correlations in terms of an AR process, and use the Variational Bayesian (VB) framework for fMRI time series (Penny et al., 2003) to estimate the optimal AR model order required for a short-TR sequence. VB implements Bayesian statistics for neuroimaging and offers a framework for parameter estimation complementary to classical statistics (Friston et al., 2002). Pertinent to the present study, the classical general linear model is extended by introducing an AR noise process of arbitrary order $p$. Instead of computing a pre-whitening matrix, serial correlations are explicitly modelled and integrated into the parameter estimation. The statistical model now reads $Y = XW + E$, where $Y$ is the $[T \times N]$ data matrix, $X$ is the $[T \times K]$ design matrix, $W$ is a $[K \times N]$ matrix of regression coefficients and $E$ is the $[T \times N]$ error matrix for $N$ voxels at $T$ points in time using $K$ regressors (Penny et al., 2007, 2005). This constitutes a spatio-temporal model of the fMRI data to directly incorporate dependencies between voxels (Penny et al., 2005). The autoregressive process at voxel $n$ is modelled as $y_n = Xw_n + e_n$, with $e_n = \tilde{E}_n a_n + z_n$. Here, $e_n$ models an AR process where $a_n$ are the AR coefficients, $\tilde{E}_n$ are the 'embedded' errors, i.e. the error of $p$ previous samples, and $z_n$ describes independent and identically distributed Gaussian errors. The priors over the regression and AR coefficients are used to model spatial dependencies across voxels, whereas the posterior factorizes over voxels and parameter types making the update equations tractable (Penny et al., 2005). Using the VB framework, model parameters are estimated by maximizing the free energy $F$, which constitutes a lower bound on the model evidence, and can be used for Bayesian model selection to obtain the optimal AR model order (Penny et al., 2003). Therefore, VB is well suited to estimate the AR coefficients and to choose an optimal noise model.

---

[1]Pre-whitening relies on accurate knowledge of the error covariance matrix $V$, which depends on a number of (scan) parameters and is therefore usually estimated from the data itself. The error covariance matrix contains the variance of the error signal with itself shifted by increasing time lag $\tau$ (Friston et al., 1994). The parameter estimation is performed in two steps. First, the residuals $r = Y - X\hat{w}$ are estimated using ordinary least squares, i.e. the pseudoinverse $X^-$ of the design matrix is used to estimate $\hat{w} = X^- Y$. The covariance matrix $V$ can then be estimated from the residuals, choosing between several parametric and non-parametric methods (Woolrich et al., 2001) with negligible bias (Marchini and Smith, 2003). Within the SPM framework (http://www.fil.ion.ucl.ac.uk/spm/), Restricted Maximum Likelihood (ReML) is used to estimate the hyperparameters $\lambda$ modelling the serial correlations. Variance estimates tend to be very noisy (Woolrich et al., 2001; Worsley et al., 2002), and it is therefore assumed that the covariance matrix $V(\lambda)$ is the same in all voxels and only the variance $\sigma^2$ differs between voxels, i.e. the pattern of serial correlations is the same in all voxels, but its amplitude is different at each voxel. This allows the pooling across voxels to give a highly precise estimate, and the error covariance matrix is then treated as a known quantity in the subsequent inference (Glaser and Friston, 2007). Since ReML requires a linear model for $V$, a first-order linear approximation of an autoregressive (AR) process of order 1 around the expansion point of $a_1 = 0.2$ is used to model the error covariance. Second, $V$ is then used to form the pre-whitening matrix $M = V^{-1/2}$ giving $MY = MXw + z$ with $z \sim N(0, \sigma^2 I)$ and $z = Me$. This model now conforms to the sphericity assumption and the degrees of freedom revert to their classical values.

As mentioned previously, physiological noise modelling can improve the validity of drawn inferences using fMRI data by reducing the required AR model order (Lund et al., 2006). Signal fluctuations related to cardiac and respiratory activity have been identified as major sources of structured noise in fMRI data (Bianciardi et al., 2009; Hutton et al., 2011). Signal fluctuations related to respiration enter the fMRI time series signal either through changes in venous oxygenation content (Windischberger et al., 2002) or through modulation of the main magnetic field (Van de Moortele et al., 2002) and higher image encoding fields (Bollmann et al., 2017; Vannesjo et al., 2015), which induce geometric distortions across the whole EPI image. Cardiac activity induces high and localized signal variability through mechanisms such as vessel pulsation (Dagli et al., 1999; Kasper et al., 2017). Physiological signal fluctuations can be modelled as a Fourier expansion of the cardiac and respiratory phase utilizing their intrinsic periodicity (Glover et al., 2000), thereby providing nuisance regressors which can then be included as covariates in the statistical analysis. Additionally, significant signal contributions related to changes in cardiac and respiratory rate have been identified causing low-frequency oscillations in fMRI time series (Birn et al., 2008, 2006; Chang et al., 2009). Another source of unwanted signal fluctuations are movement related effects (Friston et al., 1996). Changes in voxel position alter the spin history, thereby inducing signal fluctuations which can last for several seconds and depend on the voxel position in previous scans. Thus, a serially correlated signal is introduced into the fMRI time series. In summary, a range of physiological processes introduce unwanted, serially correlated signals that need to be taken into account in the modelling of fMRI time series data. Here, we investigate the impact of physiological noise modelling (including movement in addition to cardiac and respiratory activity) on the noise structure as well as on the optimal AR model order.

## Materials and Methods

### Data Acquisition

MRI data were acquired on a MAGNETOM 7T whole-body scanner (Siemens Healthcare, Erlangen, Germany) with a 32-channel head coil (Nova Medical, Wilmington, US). To achieve sufficiently short TRs, SMS EPI (Feinberg et al., 2010; Setsompop et al., 2012) was utilized to acquire multiple slices at once. The CMRR SMS implementation (release 11a) (https://www.cmrr.umn.edu/multiband/) was used and reconstruction was performed using the slice-GRAPPA technique (Setsompop et al., 2012). FMRI data were acquired with a short-TR sequence to investigate serial correlations and their interaction with physiological noise modelling. The results were then compared to those obtained from data acquired with a longer, more common TR. Imaging parameters for the short-TR sequence were: TR = 589 ms, voxel size = 2.5 mm isotropic, TE = 23 ms, SMS-acceleration-factor = 4, GRAPPA-factor = 2, FOV = 212 mm x 212 mm, number of slices = 48, number of scans = 581. Imaging parameters for the long-TR sequence were: TR = 1990 ms, voxel size = 1.3 mm isotropic, TE = 25 ms, SMS-acceleration-factor = 3, GRAPPA-factor = 3, Partial Fourier = 7/8, FOV = 212 mm x 212 mm, number of slices = 96, number of scans = 188.

### Study Design

The project was approved by the University of Queensland's Medical Research Ethics Committee. N = 10 healthy, right handed participants with normal or corrected to normal vision participated in the study after giving written, informed consent. The data were acquired as part of a larger study comparing different sequence parameter settings for fMRI at 7T. For six participants (five female), cardiac and respiratory data were recorded concurrently with the image acquisition using a breathing belt and an ECG system (Brain Products, Gilching, Germany) and their data were analysed here.

Participants performed a finger tapping task consisting of blocks of simple movement, complex movement and rest. In the simple movement condition, a visual stimulus of four dashed lines was presented and participants were asked to respond with consistent, medium length button presses of the index, middle, ring and little fingers of the right hand in sequential order. In the complex movement condition, a visual stimulus consisting of two dots and two dashed lines, indicating short and long button presses, was presented. In each block, the visual cue (2500 ms duration) followed by a fixation cross (500 ms duration) was shown six times, resulting in an 18 s block length. In the complex condition, each of the six possible combinations of two short and two long button presses were presented in a randomized order. In the rest condition, only the fixation cross was shown. For two participants, the first run of each sequence followed the order [rest - complex - simple - rest - simple - complex - repeated], while for the remaining four participants, the simple condition was presented first [rest - simple - complex - rest - complex - simple - repeated]. In total, one run of the task lasted 342 s, containing 6 blocks of rest, simple and complex movement and an additional rest block at the end. For each sequence parameter setting, two runs of fMRI data were acquired. For the second run of each sequence, the order of movement conditions was reversed.

### Preprocessing

Preprocessing was performed using SPM12 (r6224, Wellcome Trust Centre for Neuroimaging, London, UK) and Matlab (R2016a, The MathWorks, Inc., Natick, MA, US). For anatomical reference, an EPI image with an isotropic voxel size of 1 mm was chosen. Upon visual inspection, it provided better alignment in distortion prone areas such as frontal regions and around the ventricles than the additionally acquired T1-weighted image following the rationale in Grabner et al. (2014). To provide a robust starting point for the image segmentation, this reference image was first coregistered to the MNI305 T1 template. Next, the image was segmented using the unified segmentation algorithm (Ashburner and Friston, 2005) to retrieve tissue probability maps (TPMs) in subject space as well as the deformations field from and to MNI and subject space. The functional data were preprocessed in the following way: Realignment using the two-pass procedure. Coregistration (including resampling using a 7th order B-spline) to the reference anatomical EPI image for the first run, coregistration to the mean image of the first run for the second run to improve between-run alignment. Last, smoothing of the functional images was performed using a Gaussian smoothing kernel with 5 mm full-width-at-half-maximum (FWHM) size. All analyses were performed in subject space.

### Physiological Noise Modelling

Peripheral ECG data were preprocessed using an in-house implementation and the fieldtrip toolbox (Oostenveld et al., 2011) for data read-in. Following the recommendations in Ritter et al. (2007), a simple gradient artefact correction was used fitting an offset and a moving average template (computed from the current, the previous and following 4 scans), followed by low-pass filtering and downsampling to 100 Hz. Preprocessed cardiac and raw breathing data were used to compute the Fourier expansion of cardiac and respiratory phase (Glover et al., 2000) as implemented in the physIO toolbox (Kasper et al., 2017) with a cardiac model order of 3, a respiratory model order of 4, and an interaction model order of 1 (Harvey et al., 2008). Changes in respiratory and cardiac rate were modelled using the respiration response function (RRF) (Birn et al., 2008) and cardiac response function (CRF) (Chang et al., 2009), respectively. Following the recommendations in Chang et al. (2009) and Falahpour et al. (2013), and to accommodate the (negative) latency of -8 s between the estimated RRF from the breath-hold experiments and the rest data found in Birn et al. (2008), specific delay values were estimated for the RRF and CRF for each individual participant. Based on the latencies reported in Birn et al. (2008), delay values of -12, -8, -4, 0, 4, 8 and 12 s were examined. The response delay for each participant was then defined based on the highest number of supra-threshold voxels obtained in an F-test for each response function using the first short-TR run of each subject. To model remaining signal fluctuations related to movement, the Volterra expansion of the realignment parameters was used (Friston et al., 1996). The derivative, square and squared derivative of the realignment parameters totalling 24 regressors were estimated for the short-TR sequence.

### Model Estimation and Analysis

Two task regressors for the simple and complex movement condition were constructed by specifying the respective onsets and duration (18s) of each block. The resulting box functions were then convolved with the canonical hemodynamic response function using the standard parameters provided in SPM12 and its temporal and dispersion derivatives.

To investigate the impact of physiological noise modelling, the estimation process was repeated with four different sets of nuisance regressors: (i) 'no phys' uses the realignment and task regressors, but no physiological noise regressors, (ii) 'RETROICOR' adds the physiological noise regressors as described in the RETROICOR model (Glover et al., 2000) to the task and realignment regressors, (iii) 'RETROICOR +

RRF + CRF' incorporates the RETROICOR regressors and RRF (Birn et al., 2008) and CRF (Chang et al., 2009), (iv) 'RETROICOR + Volterra' contains RETROICOR regressors in combination with the Volterra expansion of the realignment parameters (Friston et al., 1996) for the short-TR sequence.

Similar to previous studies, the short-TR sequence was downsampled by a factor of 4 to create an artificial long-TR sequence (Boyacioğlu et al., 2015; Todd et al., 2017). Thereby, an fMRI time series that matches the spatial resolution of the short-TR sequence but has the temporal characteristics as if it were acquired with a TR of 2356 ms was obtained. This permitted the effect of serial correlations to be investigated on a downsampled short-TR sequence with otherwise identical signal properties compared to the short-TR sequence and on the long-TR sequence in a more realistic setting when taking full advantage of current imaging capabilities.

Bayesian model estimation was then performed to investigate the strength and characteristics of serial correlations in the fMRI time series. To this end, the VB framework (Penny et al., 2003, 2005) was used to compute log model evidence maps, containing the contribution to the overall model evidence from each voxel (Penny et al., 2007), for different AR model orders ranging from 1 to 10. This enabled the determination of the optimal AR model order in each voxel required to model the serial correlations in the fMRI data. In addition, the impact of physiological noise modelling on serial correlations was investigated by using the four different sets of nuisance regressors described above. An uninformative (flat) signal prior and unweighted graph Laplacian (Harrison et al., 2008, 2007) noise priors were chosen for the Bayesian model estimation which was performed in a slice-by-slice fashion. To include the same data for different AR model orders, and thereby to allow a comparison between log model evidence maps, the first $p-1$ scans with $p$ being the respective AR model order were removed from the data (Penny et al., 2003). The optimal AR model order for every voxel was determined as the AR model with the highest log model evidence for each set of nuisance regressors. To examine the evidence for a higher order AR-model in the presence of physiological noise modelling, log Bayes factors comparing AR model orders 4 to 1 were computed as the difference of the respective log model evidences (Kass and Raftery, 1995) for the 'RETROICOR + RRF + CRF' model for all three sequences and, in addition, the 'RETROICOR+Volterra' model for the case of the short-TR sequence.

The optimal AR model order was then summarized by computing the mean distribution of the voxel count over all subjects and runs in 6 different regions of interest (ROIs). The voxel count (in %) represents the number of voxel with AR model order 1 to 10 being the winning model, i.e. having the highest log model evidence, divided by the total number of voxel in each region. The 6 ROIs comprised the three tissue classes cerebrospinal fluid (CSF), grey matter (GM) and white matter (WM) as well as three regions expected to be involved in the task, i.e. primary motor cortex (M1), supplementary motor area (SMA) and putamen (Bednark et al., 2015). The cortical ROIs M1 and SMA were defined using the Harvard-Oxford cortical structural atlas as distributed with the FMRIB Software Library (Desikan et al., 2006; Frazier et al., 2005; Goldstein et al., 2007; Makris et al., 2006). The ROI for the putamen was derived from high-resolution 7T imaging (Keuken et al., 2014; Tziortzi et al., 2011). The ROIs were limited to grey matter by multiplying each ROI with a grey matter mask derived by thresholding the grey matter TPM at 0.9.

To visualize the noise characteristics fitted by the different AR models, their power spectra were estimated using the SPM function spm_ar_freq.m. For each voxel, the AR coefficients of the winning model were extracted and the spectra estimated. For each AR model order, the obtained spectra were then averaged and compared across the four different noise modelling schemes. Additionally, the spectrum when using an AR model of order 1 was compared to the spectrum using the optimal AR model.

Classical model estimation was performed to investigate the spectrum of the residuals visualizing the impact of physiological noise modelling and pre-whitening on the frequency content of the image time series and providing insight into possible remaining noise sources. Therein, the two different pre-whitening strategies as provided in SPM 12 were used: either the AR(1) or the FAST model, which uses a dictionary of covariance components based upon exponential covariance functions. Likewise as for the Bayesian model estimation, classical model estimation was performed for the four different noise modelling schemes with either of the pre-whitening options. The average amplitude spectrum of all grey matter voxels was then computed from the residual time series obtained after model fitting.

In summary, Bayesian model estimation was performed using data with three TRs (589 ms, 1990 ms and 2356 ms), and four noise modelling methods (no phys; RETROICOR; RETROICOR + RRF + CRF, RETROICOR + Volterra). Similarly, classical model estimation was performed using either an AR(1) model or the FAST option in SPM for pre-whitening. To investigate the effect of spatial smoothing, the analysis was repeated on the unsmoothed data obtained with the short-TR sequence.

# Results

## Optimal AR Model Orders and the Impact of Physiological Noise Modelling

To investigate serial correlations and the effect of physiological noise modelling for the short-TR sequence, optimal AR model orders obtained for the four noise modelling schemes are illustrated in Figure 1. A high optimal AR model order of up to 10 was found in large areas, especially in CSF and grey matter regions, when no physiological noise modelling was performed (Figure 1, 1st column). This indicates a complex, non-white noise structure in the fMRI time series. Including RETROICOR regressors to model cardiac and respiratory signal fluctuations successfully reduced the optimal AR model order to approximately 4 and below (Figure 1, 2nd column). High AR model orders remained in ventricles and anterior and posterior tissue-air boundaries. Additional regressors for variations in cardiac rate and respiration ('RETROICOR + RRF + CRF') had a localized effect (white arrow in Figure 1, 3rd column). Including additional movement regressors ('RETROICOR + Volterra') mainly reduced optimal AR model orders at the anterior and posterior tissue-air boundaries (white arrows in Figure 1, 4th column). In general, white matter areas showed lower AR model orders, with some voxel having an optimal AR model order of 1.

The distribution of optimal AR model orders shows the impact of physiological noise modelling for different tissue classes and in different cortical and subcortical regions (Figure 2): Grey matter had a large number of voxels with a high optimal AR model order without physiological noise modelling, i.e. 37 % of the voxels had an optimal AR model order > 5 (Figure 2, top left). For comparison, 48 % of the voxels in CSF (Figure 2, centre left), but only 20 % of the voxels in white matter had an optimal AR model order > 5 (Figure 2, top bottom). Including RETROICOR regressors had the largest impact reducing the number of voxels with optimal AR model order > 5 to 16 % in grey matter, 21 % in CSF, and 10 % in white matter. Adding RRF + CRF regressors further reduced this amount by 1 % in all three tissue classes. Additional movement regressors had a slightly larger impact, reducing the number of voxels with optimal AR model orders > 5 to 14 % in grey matter, 17 % in CSF and 9 % in white matter. In total, 68 % of the grey matter voxels had an AR optimal model order ranging between 2 and 4 when using the RETROICOR noise modelling scheme (Figure 2, top left). The highest voxel count in grey matter was obtained for an AR model order of 4, with 28 % favouring this AR model under the RETROICOR + Volterra noise modelling scheme. Interestingly, the voxel count for an optimal AR model order of 1 was < 7 % in all three tissue classes even with physiological noise modelling.

The grey matter voxels in M1 and SMA showed comparable properties with regard to voxel count and impact of physiological noise modelling as the whole grey matter. When including RETROICOR regressors 69 % of the voxels in M1 supported an optimal AR model order between 2 and 4, and only 15 % supported an AR model order > 5 (Figure 2, top right). Similarly in SMA, 60 % of the voxels supported an optimal AR model order between 2 and 4 and only 12 % supported an AR model order > 5. (Figure 2, centre right). However, the voxel count for an optimal AR model order of 1 (20 %) was much higher than in the previously discussed ROIs. Including additional regressors for cardiac and respiration response modelling (RETROICOR + RRF + CRF) or additional movement regressors (RETROICOR + Volterra) had a smaller impact giving results comparable to the RETROICOR noise model. The distribution of optimal AR model orders in the putamen showed a different pattern, with a maximum of 46 % supporting an AR model order of 2 and only 9 % of the voxels with an optimal AR model order > 3 when including the RETROICOR regressors (Figure 2, bottom right). As in the SMA, the voxel count supporting an AR model order of 1 (33 %) was much higher than in the whole grey matter.

In summary, tissue-class and region-specific distributions of optimal AR model orders were found. Including physiological noise regressors successfully reduced the number of voxel with high AR model orders and, consequently, increased the voxel count with low optimal AR model orders. Nevertheless, an AR model of order 1 proved insufficient for most voxels. Most grey matter areas still had an optimal AR model order ranging between 2 and 4.

In comparison, optimal AR model orders for the long-TR sequence remained low, with the majority of voxels having an optimal AR model order of 1 or 2 even without physiological noise modelling (Figure 3A, left column). Elevated AR model orders were observed in the vicinity of the circle of Willis and the ventricles, but also close to the insula and the anterior cingulate cortex (white arrows in Figure 3A, left column). Including regressors for physiological noise successfully reduced the optimal AR model order in these areas (Figure 3A, centre and right column), with virtually no difference between RETROICOR regressors only and the 'RETROICOR + CRF + RRF' noise modelling scheme. Optimal AR model orders for the downsampled short-TR sequence (Figure 3B) exhibited nearly identical characteristics as for the long-TR data. Slightly higher AR model orders were observed in a small number of voxels without physiological noise modelling (Figure 3B, left column). When including physiological noise regressors, low optimal AR model orders of 1 or 2 were obtained (Figure 3B, centre and right column).

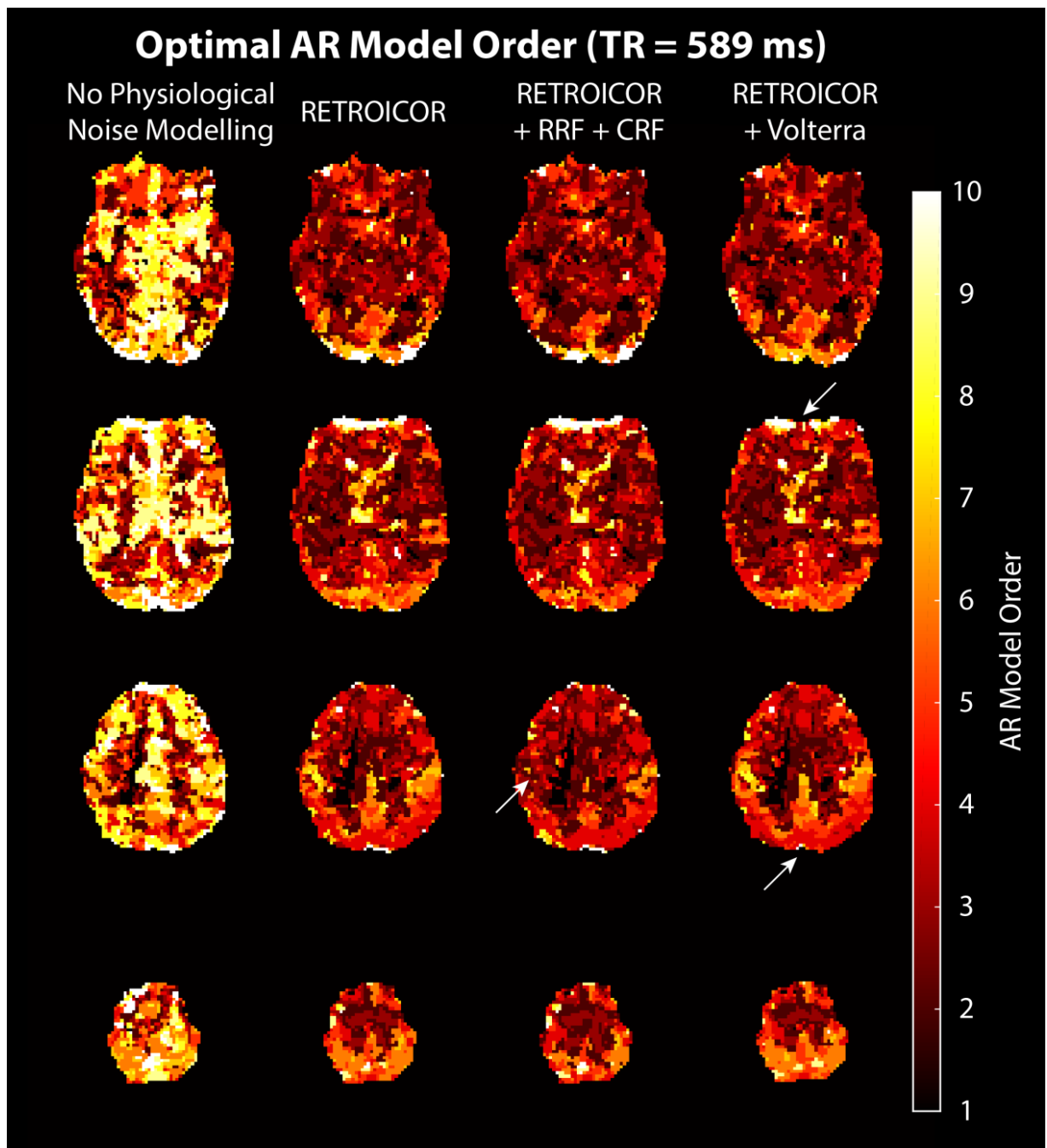

**Figure 1**: Estimated optimal AR model orders for the short-TR sequence without physiological noise modelling (1ST COLUMN), with RETROICOR regressors (2ND COLUMN), with RETROICOR regressors and cardiac and respiration response function modelling (3RD COLUMN) and with RETROICOR regressors and Volterra expansion of the realignment parameters (4TH COLUMN) illustrated on the example of 4 axial slices (subject 1, run 1, TR = 589 ms, 2.5×2.5×2.5 mm$^3$, smoothed with a 5 mm FHWM Gaussian kernel). The white arrows indicate a local reduction of optimal AR model orders for the RETROICOR + RRF + CRF noise model (3RD COLUMN) and a reduction of optimal AR model orders at anterior and posterior tissue-air boundaries for the RETROICOR + Volterra noise model (4TH COLUMN).

The voxel count across subjects and runs for the long-TR sequences showed a majority of voxel with an optimal AR model order of 1 (Figure 4A). Even without physiological noise modelling, 45 % of the grey matter (Figure 4A, top left), 39 % of the CSF (Figure 4A, centre left), and 74 % of the white matter voxels (Figure 4A, bottom left) had an optimal AR model order of 1. These numbers were increased to 52% for grey matter and CSF, and 81 % for white matter voxels when including 'RETROICOR + CRF + RRF' regressors. Accordingly, the number of voxels with optimal AR model order > 3 was below 6 % in all three tissue classes. The distribution of optimal AR model orders in M1 and SMA grey matter voxels mimicked the characteristics observed in the whole grey matter (Figure 4A, top and centre right).

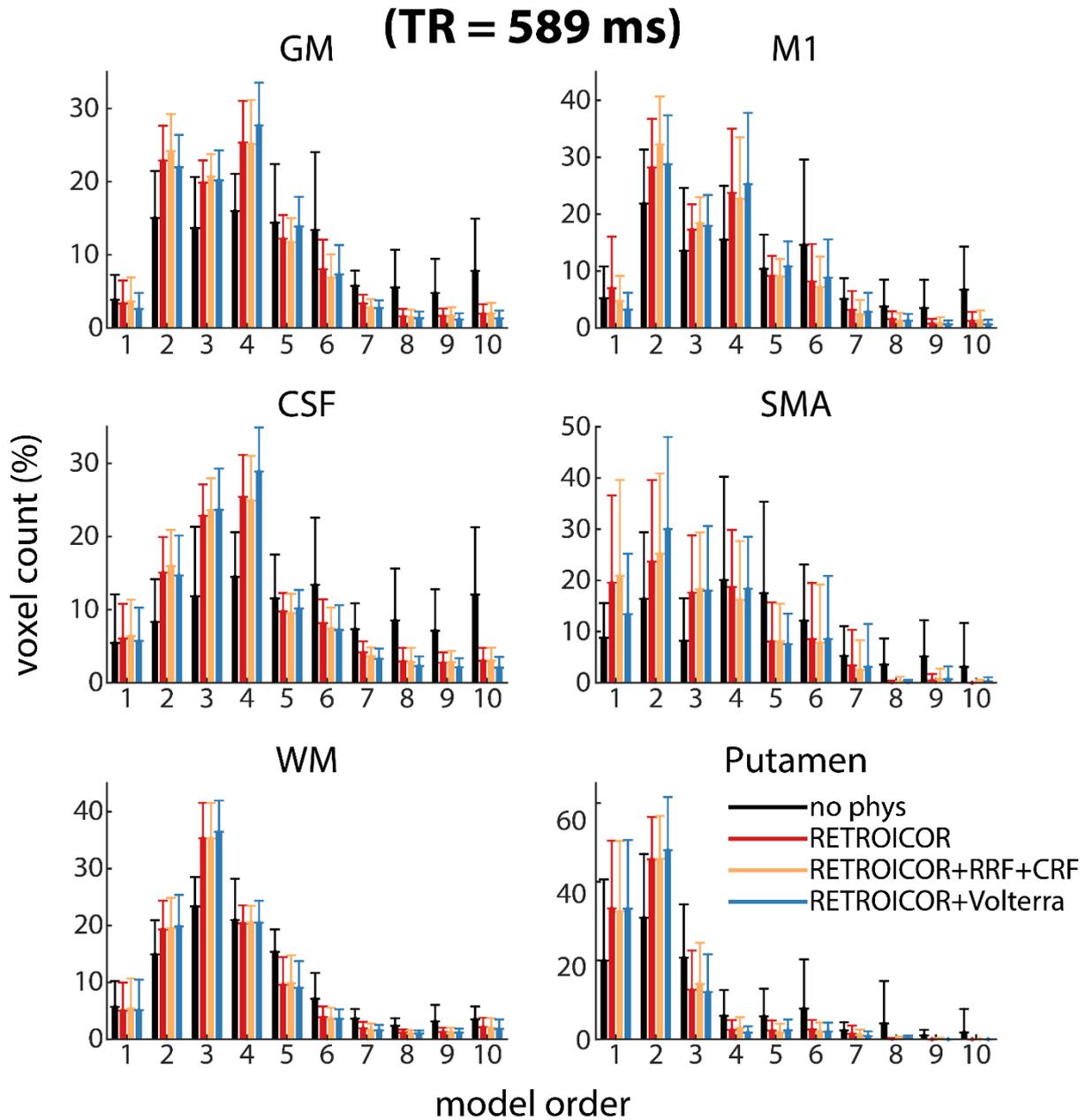

**Figure 2**: Mean and standard deviation across subjects and runs of the voxel count (%) for each optimal AR model order in 6 different regions-of-interest without physiological noise modelling (black), with RETROICOR regressors (red), with RETROICOR regressors and cardiac and respiration response function modelling (orange) and with RETROICOR regressors and Volterra expansion of the realignment parameters (blue) for the short-TR sequence (TR = 589 ms, 2.5×2.5×2.5 mm³, smoothed with a 5 mm FHWM Gaussian kernel). The voxel count represents the number of voxel with AR model order 1 to 10 being the winning model, i.e. having the highest log model evidence, divided by the total number of voxel in each region.

As observed for the short-TR sequence, the distribution in the putamen is distinct from the previous ROIs and resembled features comparable to white matter voxels with 85 % of the voxels having an optimal AR model of order 1 when including RETROICOR + RRF + CRF regressors (Figure 4A, bottom right). This number was only slightly reduced (to 80 %) when no physiological noise modelling was performed. Optimal AR model orders for the downsampled short-TR sequence follow a similar distribution, but with slightly higher values for AR model order 2 (Figure 4B). Including RETROICOR + RRF + CRF regressors, 49 % of the voxel in CSF (Figure 4B, top left), 51 % of the voxel in grey matter (Figure 4B, centre left) and 70 % of the voxel in white matter (Figure 4B, bottom left) had an optimal AR model order of 1. The number of voxels with optimal AR model order > 3 was below 3 % in all three tissue classes. Again, M1 (Figure 4B, top right) and SMA (Figure 4B, centre right) had comparable properties as the whole grey matter, with 51 % and 43 % of the voxel having an optimal AR model order of 1. Similarly, the highest number of voxels with optimal AR model order 1 (80 %) were found in the putamen.

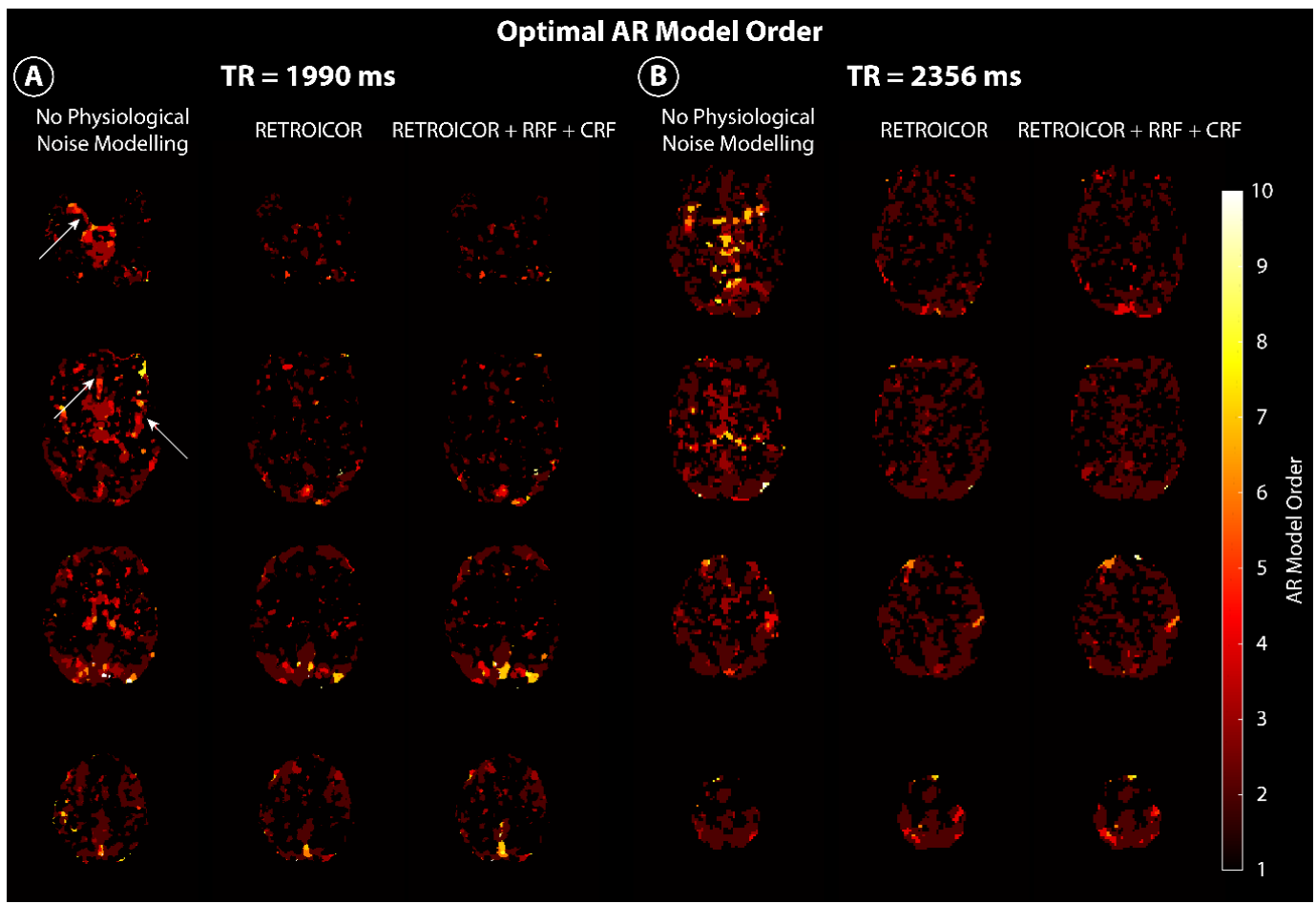

**Figure 3**: Estimated optimal AR model orders without physiological noise modelling (Left), with RETROICOR regressors (Centre) and with RETROICOR regressors and cardiac and respiration response function modelling (Right) for (A) the long-TR sequence (subject 1, run 1, TR = 1990 ms, 1.3×1.3×1.3 mm³, smoothed with a 5 mm FWHM Gaussian kernel) and (B) the downsampled short-TR sequence (subject 1, run 1, TR = 2356 ms, 2.5×2.5×2.5 mm³, smoothed with a 5 mm FWHM Gaussian kernel). The white arrows indicate elevated AR model orders obtained without physiological noise modelling in the vicinity of the circle of Willis, the insula and the anterior cingulate cortex.

In summary, the choice of the physiological noise modelling scheme has limited influence on the optimal AR model order for the long-TR sequence and the downsampled short-TR sequence, with largest effects in CSF and grey matter regions. A majority of voxel have an optimal AR model order of 1 or 2 even in the absence of physiological noise modelling.

### Bayes Factor Analysis

To assess the statistical significance of the obtained optimal orders, log Bayes factors were computed as the difference in log model evidence (Kass and Raftery, 1995). For the short-TR sequence, log Bayes factors comparing an AR model of order 4 to an AR model of order 1 showed positive evidence (BF > 3, corresponding to a posterior model probability > 95 %) for the higher AR model order in large areas of the brain, even when including RETROICOR + RRF + CRF (Figure 5, 1st column) or RETRICOR + Volterra regressors (Figure 5, 2nd column). Some white matter voxel showed support for the AR(1)-model as expected from Figure 1, where AR(1) was the winning model.

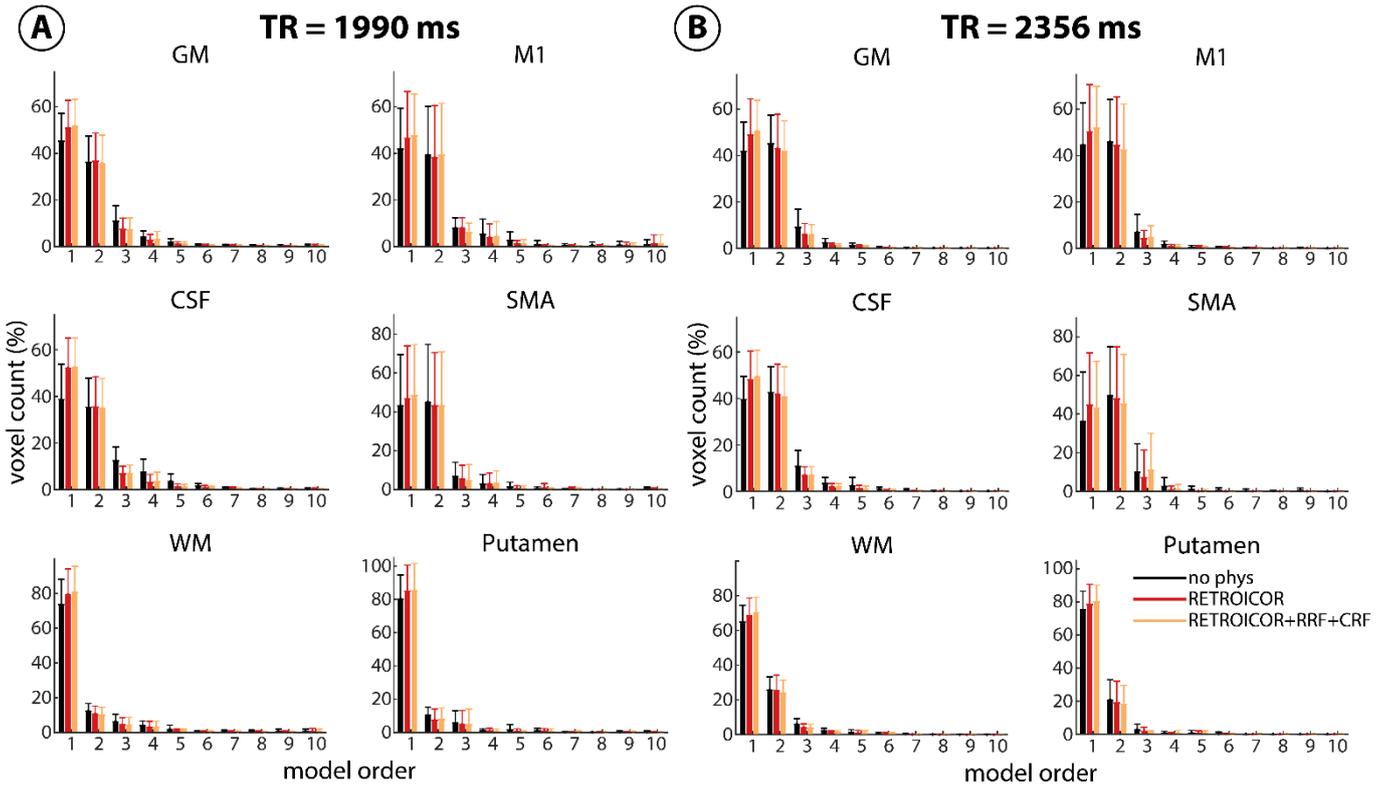

**Figure 4**: Mean and standard deviation across subjects and runs of the voxel count (%) for each optimal AR model order in 6 different regions-of-interest without physiological noise modelling (black), with RETROICOR regressors (red), and with RETROICOR regressors and cardiac and respiration response function modelling (orange) for (A) the long-TR sequence (TR = 1990 ms, 1.3×1.3×1.3 mm$^3$, smoothed with a 5 mm FWHM Gaussian kernel) and (B) the downsampled short-TR sequence (TR = 2356 ms, 2.5×2.5×2.5 mm$^3$, smoothed with a 5 mm FWHM Gaussian kernel). The voxel count represents the number of voxels with AR model order 1 to 10 being the winning model, i.e. having the highest log model evidence, divided by the total number of voxels in each region.

In contrast, for the long-TR sequence positive evidence was mostly found for the AR(1)-model, with only a few patches favouring an AR model order of 4 (Figure 5, 3rd column). Similarly for the downsampled short-TR sequence, positive evidence was mostly found for the AR(1) model compared to the AR(4) model (Figure 5, 4th column). However, in large areas of the brain, log Bayes factors remain very low, indicating no clear evidence for any of the two models.

### AR Coefficient Analysis

The spectra in Figure 6 illustrate the noise characteristics that were fitted by the different AR models. Frequencies found in the fMRI time series are down weighted, i.e. showing a dip in the estimated spectra. For example, the spectra for AR model order 10 without physiological noise modelling (Figure 6, 2nd column, bottom) showed a clear dip at the cardiac frequency (∼ 0.66 Hz) indicating that a strong signal was present at this frequency and has been fitted by the AR model. Across all AR model orders, damping of low frequency oscillations was observed (Figure 6, left). Up to AR model order 4, no differences in the spectra obtained from the different noise modelling schemes were found. However, the spectra showed increased damping of low frequency components with increasing AR model order. For even higher AR model orders, cardiac and respiratory frequencies can be found in the spectrum, especially in the case of no physiological noise modelling. Comparing the spectrum of the winning model with the AR(1)-model when including RETOICOR, RRF and CRF regressors shows that the AR(1)-model can only approximate the required, more complex shape of the power spectra in the voxel with higher optimal AR model order (Figure 6, right). The largest differences were observed for the high frequency part of the spectrum.

The estimated AR parameter maps retain anatomical structures indicative of tissue specific noise processes (Figure 7). High first order AR coefficients > 0.8 were obtained for the AR(1) and the AR(4) model in grey matter regions. Higher order AR coefficients were mainly present in grey matter voxels and nearly zero in white matter (Figure 7, bottom left). Including RETROICOR regressors reduced the higher order coefficients values, providing a more homogenous spatial distribution (Figure 7, bottom right).

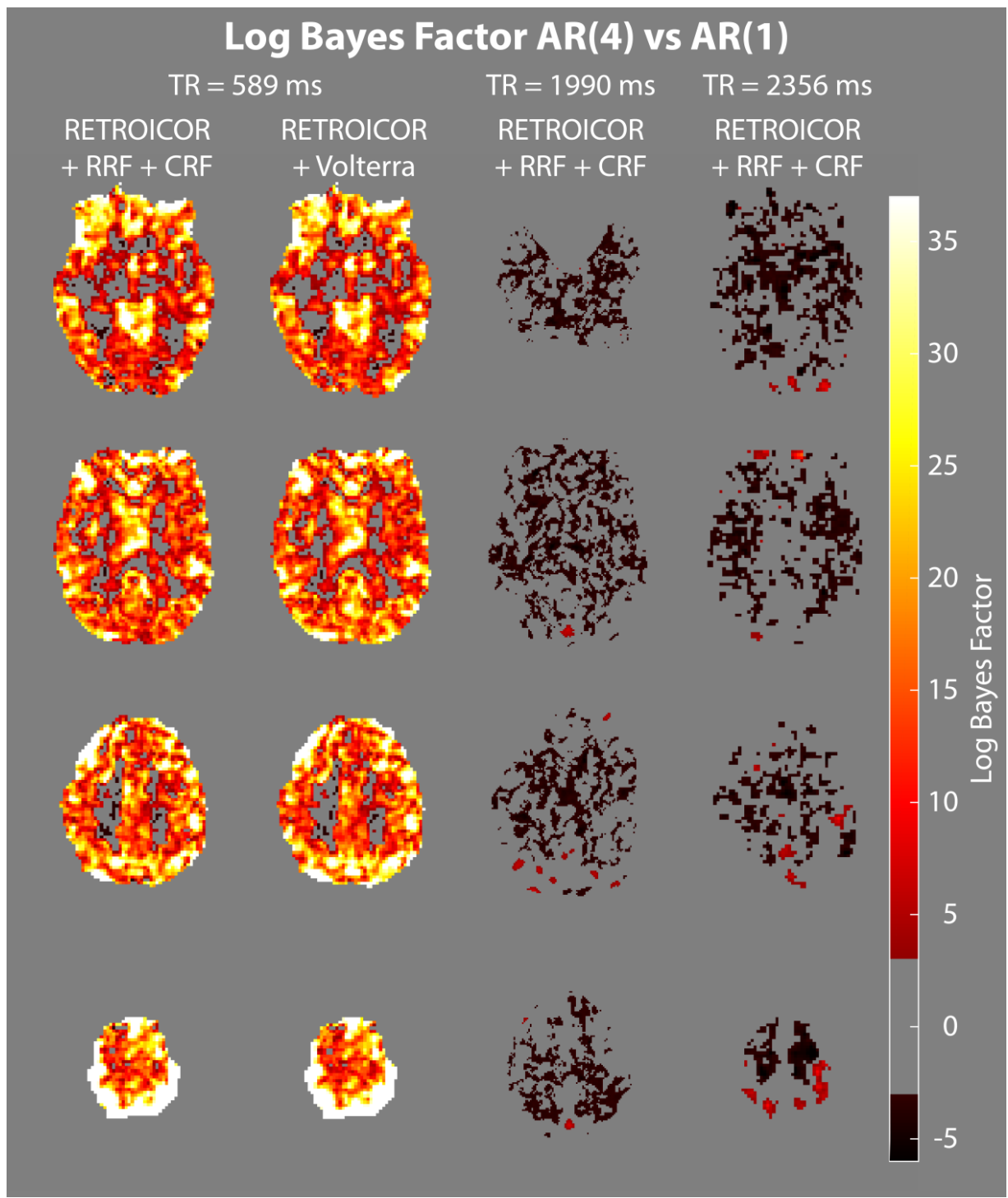

**Figure 5**: Log Bayes Factor for AR model order 4 vs. AR model order 1 for the short-TR sequence with RETROICOR regressors and cardiac and respiration response function modelling (1ST COLUMN), RETROICOR regressors and Volterra expansion of the realignment parameters (2ND COLUMN), the long-TR sequence with RETROICOR and cardiac and respiration response modelling (3RD COLUMN) and the downsampled short-TR sequence with RETROICOR and cardiac and respiration response modelling (4TH COLUMN) (subject 1, run 1, smoothed with a 5 mm FWHM Gaussian kernel). High log Bayes factors (> 3) indicate positive evidence for an AR model of order 4, whereas negative log Bayes factors below -3 indicate support for an AR model of order 1. Log Bayes factors between -3 and 3 indicate no clear evidence for either model.

### Noise Spectrum and FAST for Pre-whitening

The average spectrum of the residual image time series visualizes remaining noise contributions after classical model estimation (Figure 8 and Figure 9). The spectrum of the 'no phys' modelling scheme was utilized to identify possible remaining noise sources. In particular, the short-TR sequence showed distinct peaks in good agreement with the measured heart and breathing rates (Figure 8): cardiac – 0.66Hz (aliased), respiration – 0.31 Hz, interaction – 0.35 Hz, as well as low frequency oscillations in grey matter following a 1/f characteristic. Including RETROICOR regressors successfully removed the cardiac and respiratory frequency components. The impact of the low frequency regressors modelling cardiac and respiratory variations (RETROICOR + RRF + CRF) or remaining movement-related signal fluctuations (RETROICOR + Volterra) was limited, with most of the non-white noise characteristics in the low frequency range still being present. Comparing the two pre-whitening options, the FAST noise model (Figure 8, bottom) reduced the remaining 1/f noise compared to the AR(1) model (Figure 8, top). Note that the cardiac and respiratory components remain in the spectrum. Adding RETROICOR regressors removed these physiological noise peaks, resulting in a nearly flat, i.e. white, noise distribution across frequencies.

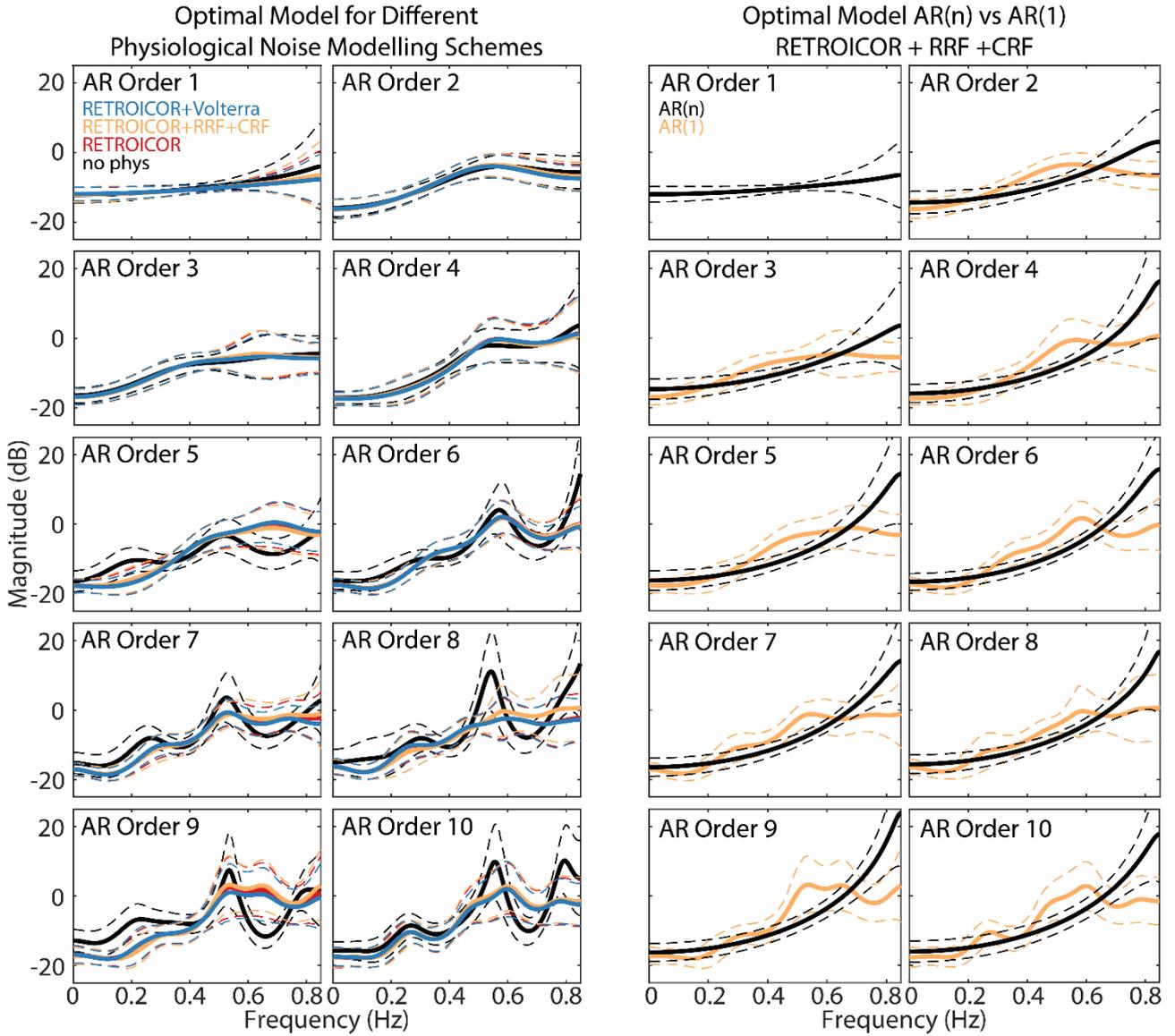

**Figure 6**: Spectrum (solid line = mean, dotted line = standard deviation) of different AR models in grey matter for the short-TR sequence (subject 1, run 1, TR = 589 ms, 2.5×2.5×2.5 mm³, smoothed with a 5 mm FWHM Gaussian kernel). Spectrum without physiological noise modelling (black), with RETROICOR regressors (red), with RETROICOR regressors and cardiac and respiration response function modelling (orange) and with RETROICOR regressors and Volterra expansion of the realignment parameters (blue) when using only AR coefficients from matching optimal AR model orders (LEFT). Spectrum of AR(1) (orange) and spectrum of matching higher order models (black) when using RETROICOR regressors and cardiac and respiration response function modelling (RIGHT).

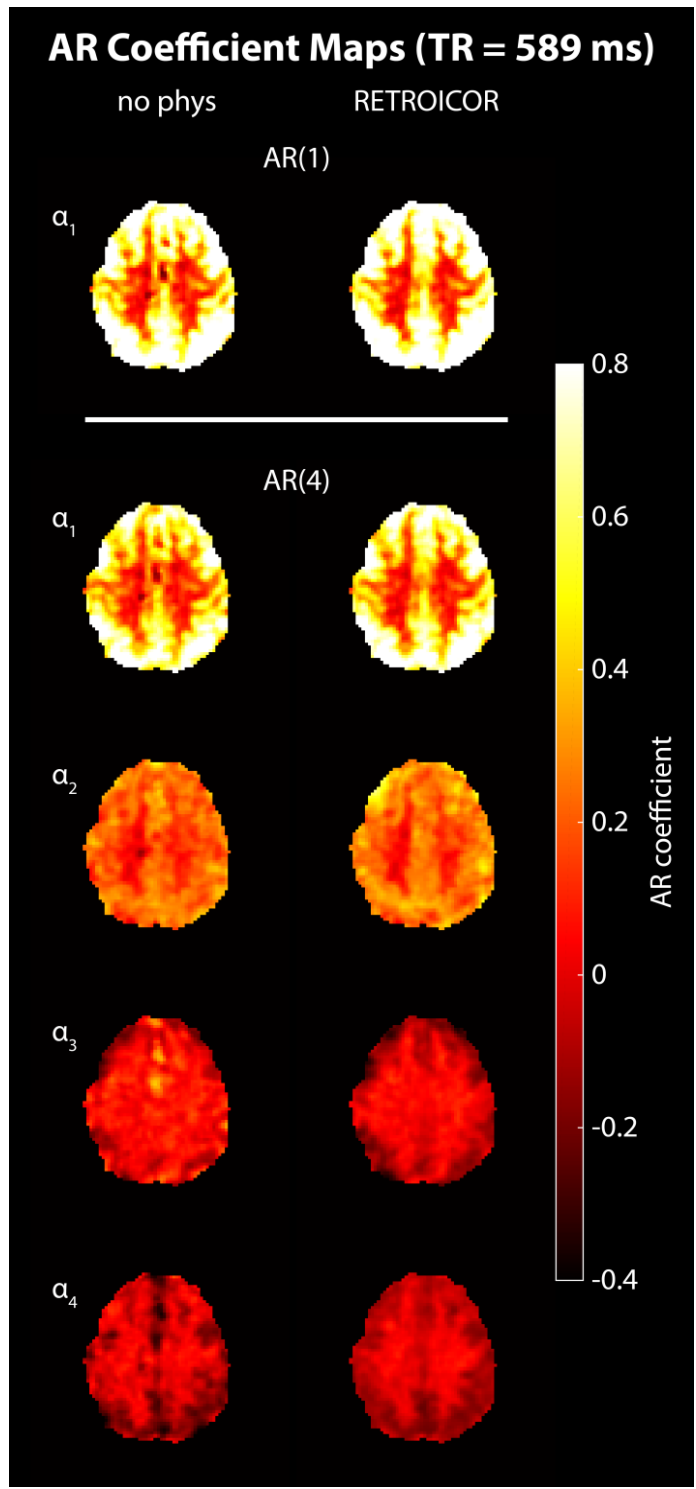

**Figure 7**: AR coefficient maps in one axial slice for an AR model of order 1 (TOP) and order 4 (BOTTOM) without physiological noise modelling (LEFT) and with the RETROICOR model (RIGHT) (subject 1, run 1, TR = 589 ms, 2.5×2.5×2.5 mm$^3$, smoothed with a 5 mm FWHM Gaussian kernel). High first order coefficients were found in all grey matter voxels for both AR model orders and physiological noise modelling approaches. Higher order coefficients retain anatomical structure up to 4$^{th}$ order (BOTTOM). However, RETROICOR modelling reduces the coefficient values successfully with a more homogenous distribution of the higher order coefficients (BOTTOM, RIGHT).

The average spectrum of the long-TR sequence showed a nearly flat noise distribution across frequencies when pre-whitening with the AR(1) model (Figure 9A, top) or FAST (Figure 9A, bottom). Individual contributions from the heavily aliased physiological frequencies (0.09 Hz – cardiac, 0.16 Hz – respiration, and 0.1 Hz – interaction) could not be discerned. Including physiological noise regressors reduced the energy content across a broad range of frequencies, with only a slight reduction in peak height and number. Similarly for the downsampled short-TR sequence, both pre-whitening options resulted in a nearly flat noise distribution (Figure 9B) with not distinct peaks at the aliased physiological frequencies (0.18 Hz – cardiac, 0.1 Hz – respiration and 0.08 Hz – interaction).

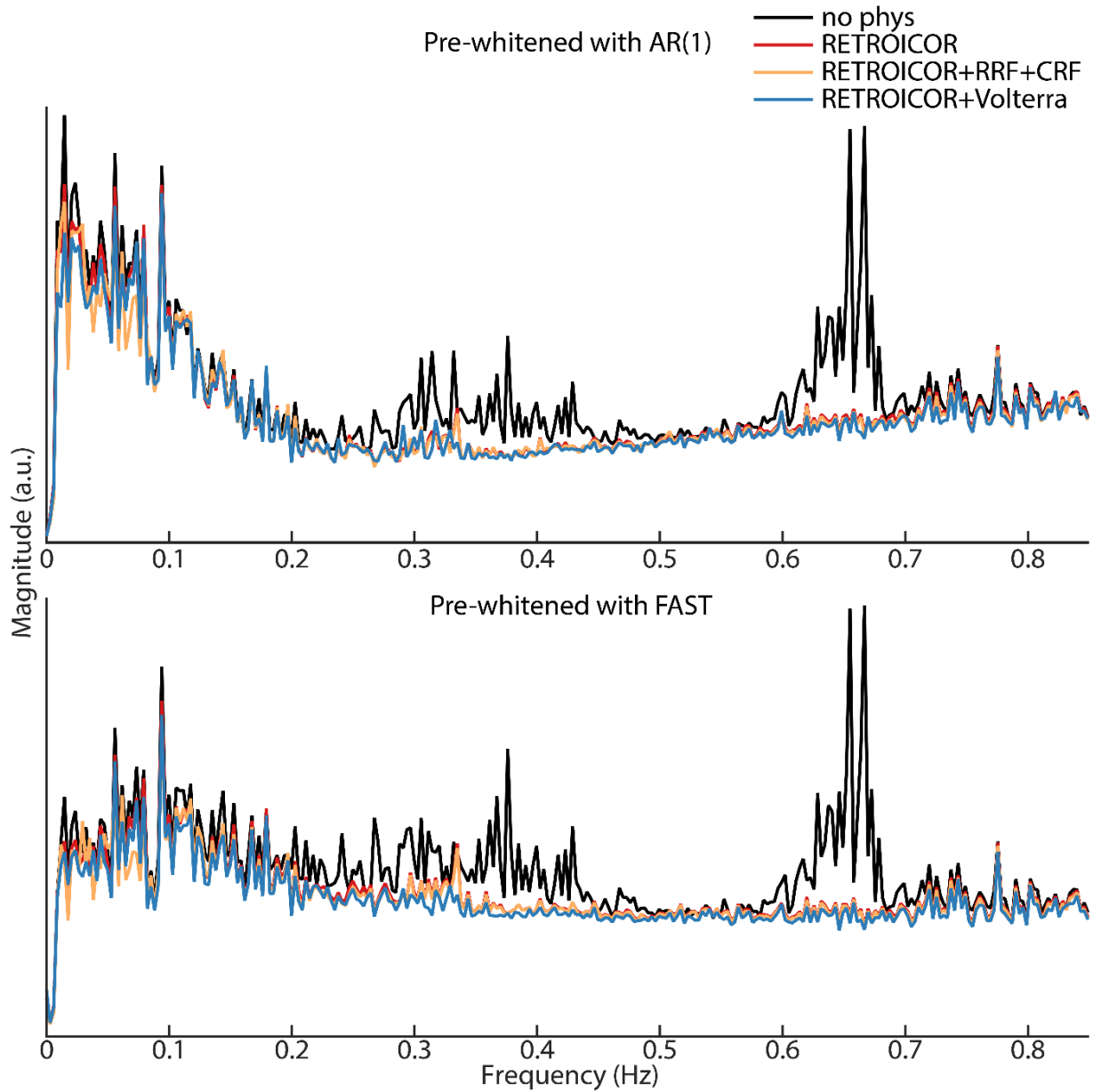

**Figure 8**: Average spectrum of the residual image time series using the classical model estimation algorithm and pre-whitening with an AR(1) model (TOP) or the FAST option (BOTTOM) without physiological noise modelling (black), with RETROICOR regressors (red), with RETROICOR regressors and cardiac and respiration response function modelling (orange) and with RETROICOR regressors and Volterra expansion of the realignment parameters (blue) for the short-TR sequence (subject 1, run 1, TR = 589 ms, 2.5×2.5×2.5 mm³, smoothed).

### Impact of Smoothing

Estimating optimal AR model orders on the un-smoothed data showed very high AR model orders mainly in CSF-bearing regions, but also close to the insula and the anterior cingulate cortex (white arrows in Figure 10, 1st column), as well as moderate optimal AR model orders in grey matter regions without physiological noise modelling (Figure 10, 1st column). Nearly all white matter voxels showed an optimal AR model order of 1. Including RETROICOR regressors successfully reduced the higher AR model orders in CSF and grey matter regions, with a remaining AR model order of 2 to 4 in most grey matter areas (Figure 10, 2nd column). Some voxels with AR model order 10 can be found in posterior and anterior regions near tissue-air boundaries (white arrow in Figure 10, 2nd column). Adding RRF + CRF regressors had a limited impact on optimal AR model orders (Figure 10, 3rd column). The additional movement regressors (RETROICOR + Volterra) reduced optimal AR model orders in a small number of voxels at the posterior and anterior tissue-air boundaries (white arrow in Figure 10, 4th column).

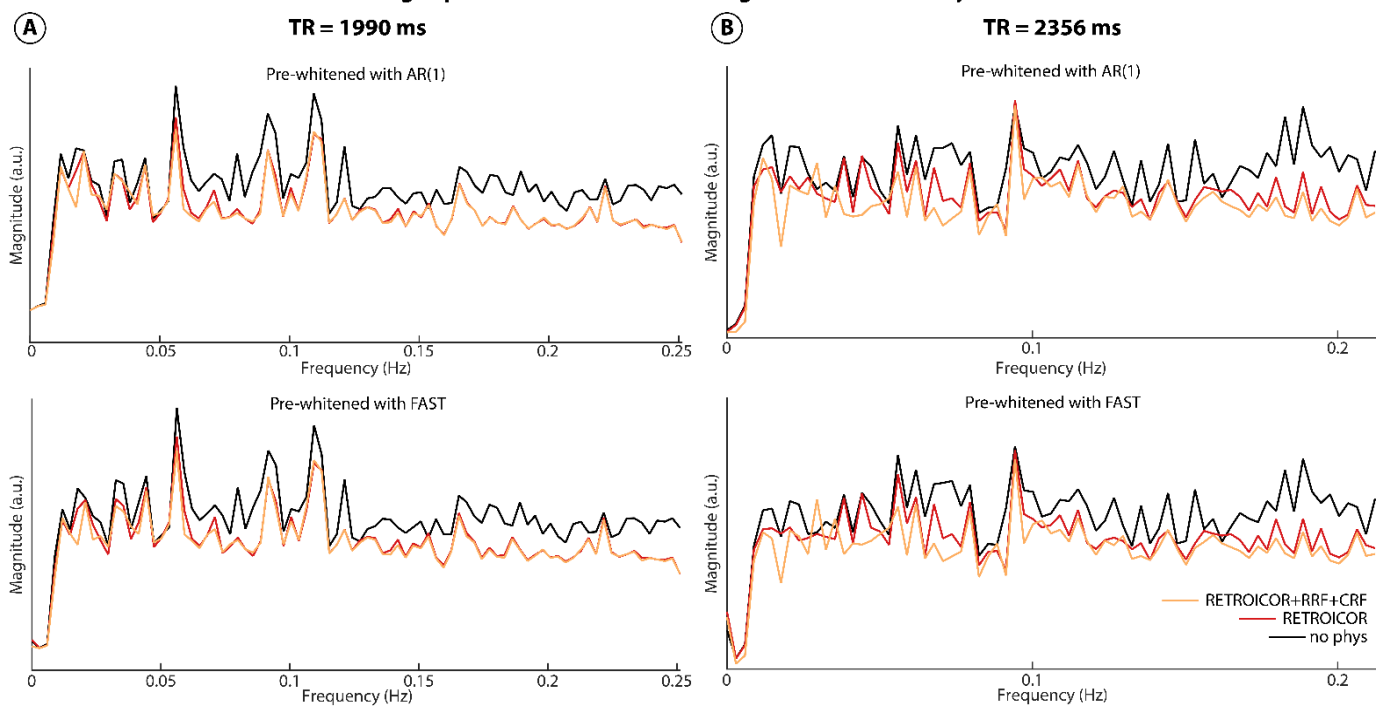

**Figure 9**: Average spectrum of the residual image time series using the classical model estimation algorithm and pre-whitening with an AR(1) model (TOP) or the FAST option (BOTTOM) without physiological noise modelling (black), with RETROICOR regressors (red) and with RETROICOR regressors and cardiac and respiration response function modelling (orange) for the long-TR sequence (A) (subject 1, run 1, TR = 1990 ms, 1.3×1.3×1.3 mm³, smoothed with a 5 mm FWHM Gaussian kernel) and the downsampled short-TR sequence (B) (subject 1, run 1, TR = 2356 ms, 2.5×2.5×2.5 mm³, smoothed with a 5 mm FWHM Gaussian kernel).

## Discussion

### Effect of TR and Physiological Noise on Optimal AR Model Order

The high AR model orders found in the short-TR sequence indicate a complex noise covariance structure that needs to be taken into account when drawing inference on single-subject fMRI data. As expected, physiological noise regressors as provided through the RETROICOR model (Glover et al., 2000) successfully reduced serial correlations in the fMRI time series, resulting in decreased optimal AR model orders. However, optimal AR model orders ranging between two and four were still found even after including physiological noise regressors.

The broad distribution of optimal AR model orders across all grey matter voxels contradicts remaining task related activity as a possible noise source. Also, the low optimal AR model order found in white matter excludes the possibility of hardware related sources which would affect grey and white matter equally. In the power spectra in Figure 6 and the residuum spectra in Figure 8 not a single unique frequency could be identified driving the observed effects. Rather, a broad 1/f noise characteristic and large variance around the cardiac peaks were observed. This indicates un-modelled neuronal activity (Bianciardi et al., 2009), given the high intrinsic serial correlation of the hemodynamic response function (Arbabshirani et al., 2014), as well as remaining physiological fluctuations not captured by the linear RETROICOR model as possible sources. Noise sources of even higher frequency seem unlikely other than the known cardiac and respiratory fluctuations and their higher harmonics (Weisskoff et al., 1993).

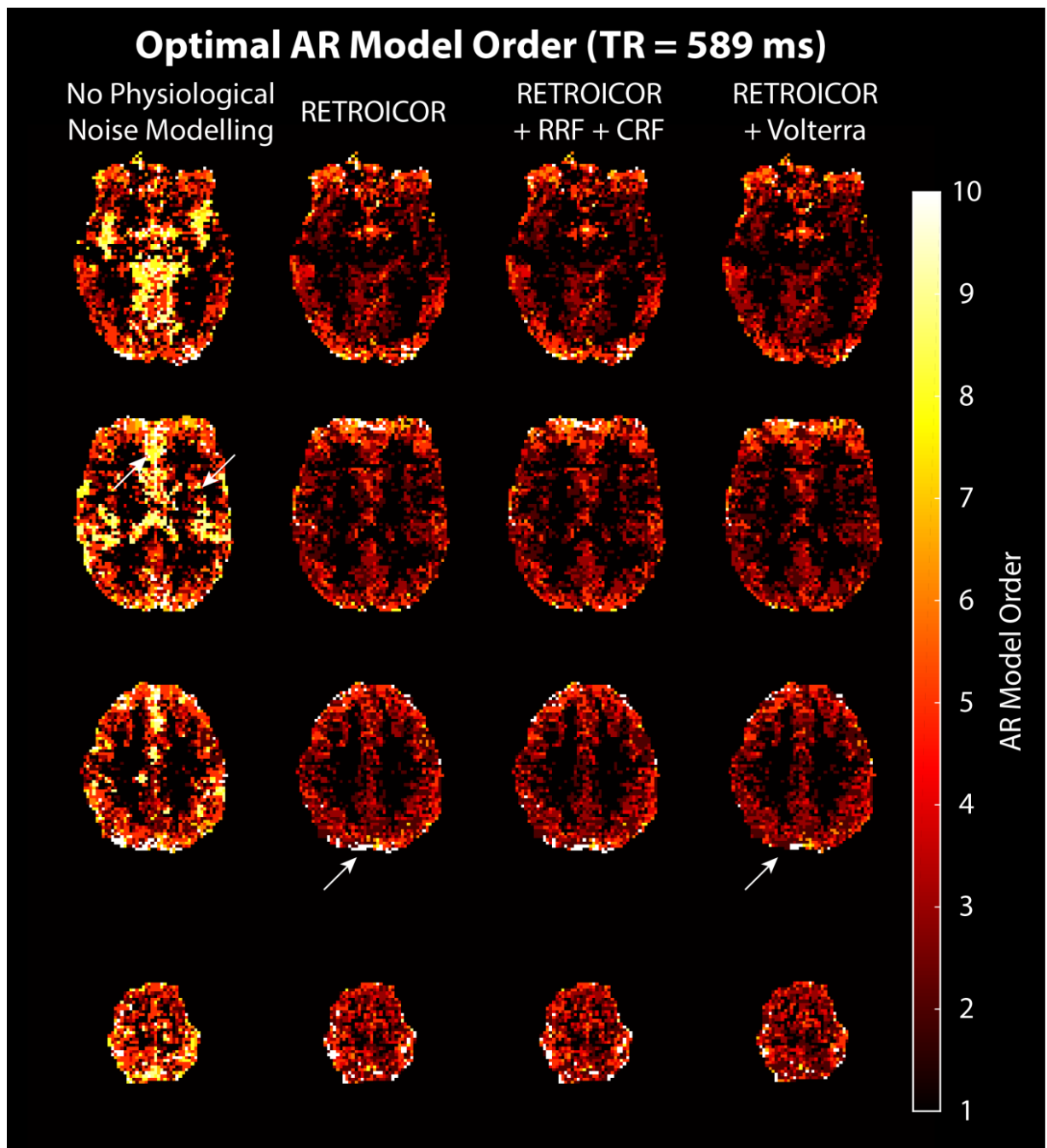

**Figure 10**: Estimated optimal AR model orders for the short-TR sequence without physiological noise modelling (1ST COLUMN), with RETROICOR regressors (2ND COLUMN), with RETROICOR regressors and cardiac and respiration response function modelling (3RD COLUMN) and with RETROICOR regressors and Volterra expansion of the realignment parameters (4TH COLUMN) when no smoothing was applied (subject 1, run 1, TR = 589 ms, 2.5×2.5×2.5 mm³). The white arrows indicate high remaining optimal AR model orders at posterior and anterior tissue-air boundaries with RETROICOR regressors (2ND COLUMN), and their slight reduction when including RETROICOR + Volterra regressors (4TH COLUMN).

Accounting for remaining low frequency oscillations through cardiac and respiration response function modelling had a comparatively limited impact, indicating either their localized reach or that more individualized response functions might be required (Falahpour et al., 2013). If specialized equipment is available, more direct measures of end-tidal CO2 through carbon-dioxide data (Wise et al., 2004) and blood flow and oxygenation using near-infrared spectroscopy data (Frederick et al., 2012) could improve the modelling of physiological processes in the low-frequency range. Including the Volterra expansion of the realignment regressors to model remaining movement related signals (Friston et al., 1996) had a small impact mainly limited to voxels at tissue-air boundaries. Importantly, motion regressors characterizing translation and rotation were included in all analysis schemes, and therefore, apparent bulk motion through respiration-induced fluctuations in the main magnetic field is included even in the 'no phys' option. Hence, serial correlations introduced through respiration could be slightly underestimated in the results presented here.

Low AR model orders were obtained for the long-TR and the downsampled short-TR sequence in line with previous studies (Lund et al., 2006; Penny et al., 2003). Despite their different spatial resolutions, virtually identical results were obtained from both sequences, indicating that extensive aliasing in sequences with a TR above 2 seconds presumably renders individual noise processes indistinguishable, resulting in low optimal AR model orders. This is in line with the observation of a negligible change in t-values with increasing AR model order for data acquired with long TRs and a reduction in t-values with increasing AR model order for data acquired with shorter TRs (Sahib et al., 2016).

### Implications for Error Modelling in Statistical Inference

When using pre-whitening to account for serial correlations, the complex noise structure found in the data acquired with sub-second TR needs to be accommodated in the estimation of the error covariance $V$. When choosing the AR model order for spectral estimation, too low AR model orders result in too smooth estimates, whereas too high AR model orders can lead to spurious peaks (Schlindwein and Evans, 1992). Applied to empirical data, using a higher AR model order rather than a too low AR model order has been recommended for spectral estimation (Boardman et al., 2002; Schlindwein and Evans, 1992). Alternatively, the FAST noise model introduced in SPM12 has shown promising results, as indicated in the frequency spectra of the residuals (Figure 8, bottom). However, the cardiac and respiratory peak could not be accounted for, and, therefore, additional noise modelling strategies targeting these specific frequencies are required. Either identifying the noise sources and incorporating nuisance regressors for physiological fluctuations or specifying a comprehensive noise model that can capture serial correlations obtained from a broad range of TR values might be the way forward.

As observed previously (Kaneoke et al., 2012), noise characteristics varied across tissue classes and cortical and subcortical regions. Grey matter voxels showed properties of both CSF and white matter, which could be partially introduced through the applied spatial smoothing and partial volume effects. Indeed, the optimal AR model orders obtained for the unsmoothed data show extended white matter areas with optimal AR model order 1, elevated AR model orders in grey matter, and extreme values almost exclusively in CSF-bearing regions. While smoothing might introduce noise in adjacent voxels, it also considerably increases the signal-to-noise ratio (SNR) and functional sensitivity. These results highlight that careful selection of voxels for noise estimation purposes is required, as previously suggested in Purdon et al. (2001) and Woolrich et al. (2001). Pooling across voxels to estimate the stationary temporal covariance $V$ might thus remain feasible when carefully selecting voxels with similar noise properties.

When using the VB framework to infer about regional activation in response to a task, spatial priors express the spatial contingency of evoked responses (Penny et al., 2005) as well as any further prior knowledge. Given the distinct noise properties of different tissue classes found in this work and previous studies (Penny et al., 2003; Woolrich et al., 2001), one might conclude that tissue specific priors could improve noise modelling. However, Penny et al. (2007) showed that Gaussian Markov Random Field priors (Woolrich et al., 2004), which assume the AR coefficient to vary smoothly, supersede tissue specific priors modelling the spatial variability in serial correlations. Given the observed variance in optimal AR model orders (Figure 10) and AR coefficient values (Figure 7) between different cortical and subcortical regions we anticipate a similar outcome for the data presented here.

### Limitations and Considerations

When employing SMS acceleration to increase BOLD sensitivity in fMRI time series, two counteracting mechanisms need to be considered. On one hand, the temporal SNR decreases with increasing SMS acceleration factor (Chen et al., 2015). On the other hand, the number of samples per unit time increases, although the effective degrees of freedom do not rise at the same rate due to serial correlations. First investigations show that moderate SMS acceleration factors between two and six strike a balance between the two (Todd et al., 2016). It is clear, however, that the effects of physiological noise needs to be further evaluated when employing SMS acceleration, given its determining role for temporal SNR (Todd et al., 2017; Triantafyllou et al., 2005). Importantly, the spatial variability of different noise processes might render different SMS-factors optimal for different ROIs (Todd et al., 2017).

We have limited our considerations to the single-subject level, keeping in mind the increased interest in using ultra-high field fMRI in single subject studies (Branco et al., 2016; De Martino et al., 2011; Stephan et al., 2017). However, group studies employing mixed-effect models also rely on the precise estimation of effect sizes and their variance (Chen et al., 2012). Since an unbiased estimator is used for the parameters, random-effects analyses are more robust against deviations in the error covariance estimation.

## Conclusion

Unlike fMRI time series with a longer, more conventional TR of ~ 2 s, SMS EPI data with a short TR of ~ 600 ms exhibit a complex noise structure that cannot be captured by an AR(1) model. While physiological noise modelling successfully reduces serial correlations, an advanced noise model is still required to account for the non-white noise content. Otherwise, single-subject analyses of fMRI data with sub-second TR will result in increased false positive rates, effectively declaring voxels as active without the prescribed significance. Hence, for valid inference on single subject-level with sub-second TR, more advanced pre-whitening schemes in combination with physiological noise modelling are necessary.

## Acknowledgements

We thank Will Penny for valuable comments on the VB algorithm and the estimation and interpretation of the frequency response of the AR models. We thank Aiman Al-Najjar, Nicole Atcheson and Jake Palmer for help with data collection. We thank the reviewers for their time and efforts and Mattias Villani and Lars Kasper for helpful comments on the pre-print. This research was supported by the National Health and Medical Research Council (APP 1088419). MB acknowledges funding from Australian Research Council Future Fellowship grant FT140100865. SB acknowledges support through the Australian Government Research Training Program Scholarship. The authors acknowledge the facilities of the National Imaging Facility (NIF) at the Centre for Advanced Imaging, University of Queensland.